\documentclass[prd,nofootinbib,preprint]{revtex4}
\usepackage[utf8]{inputenc}
\usepackage{amsmath}
\usepackage{amssymb}
\usepackage{textcomp}
\usepackage{array,multirow}
\usepackage{slashed}
\usepackage{tabularx}
\usepackage{hyperref}
\usepackage{graphicx}
\usepackage{caption}
\usepackage{xcolor}
\usepackage{textcomp}
\usepackage{cancel}
\usepackage[labelformat=parens]{subcaption}
\usepackage[toc,page]{appendix}
\usepackage{varwidth}
\begin{document}
\title{Neutrino mass observables and non-Hermitian version of Type-I seesaw model }
\author{Sasmita Mishra}
\email{mishras@nitrkl.ac.in}
\affiliation{Department of Physics and Astronomy, National Institute of Technology Rourkela, Sundargarh, Odisha, India, 769008}

\begin{abstract}
We study the non-Hermitian extension of the Lagrangian of the Standard Model extended
by singlet right-handed heavy neutrinos. The Yukawa coupling matrices comprise of hermitian and non-hermitian components and the neutrino mass eigenvalues are calculated
for three generation case. The increased number of unknown parameters in the theory due to non-hermitian nature of coupling matrices impose problem for its productiveness. Hence we apply four
zero texture for the Yukawa matrices. We consider the normal hierarchy of neutrino mass ordering
and find the mixing angles and Dirac CP violating phase by diagonalizing the lepton mass matrices. The parameters of the mixing matrix are obtained in terms of the parameters of the
non-hermitian Yukawa coupling matrix. The values of the angles and phase are compatible with experimental data available for neutrino mass observables and using this  the non-hermitian parameters are constrained. Also, we study the 
constraints imposed by $CP$ violation in the lepton sector and leptogenesis
on the model parameters.
\end{abstract}

\maketitle
\maketitle
\flushbottom

\section{Introduction}
The experiments based on neutrino oscillation phenomenon have established the fact
that this abundant particle in the Universe has non-zero mass
\cite{Fukuda:1998mi},\cite{Ahmad:2002jz},\cite{Ahmad:2002ka},\cite{Bahcall:2004mz}. The allowed ranges of neutrino oscillation parameters; three mixing angles, $\theta_{12}$, $\theta_{23}$, $\theta_{13}$, two mass squared differences $\Delta m^{2}_{21}$, $|\Delta m_{31}^{2}|$ ($\Delta m^{2}_{ij} = m_i^2 -m_j^2$) and one Dirac CP-violating phase $\delta$ by latest neutrino oscillation experiments are given in table-\ref{neop} \cite{Esteban:2020cvm}. This is in contrast
to the prediction of the standard model (SM), which otherwise is a very successful
theory of particle physics at low energy. Subsequently one relies on theories beyond
the SM in order to circumvent the problem. The canonical or Type - I seesaw mechanism
\cite{Minkowski:1977sc},\cite{Yanagida:1979as},\cite{Mohapatra:1979ia}  is one such leading
candidate which accounts for non-zero neutrino mass by adding the extra heavy right-handed singlet neutrinos to the SM.
In addition, the said mechanism can contribute to the baryon asymmetry of the Universe (BAU)
through the out of equilibrium decays of right-handed neutrinos through leptogenesis mechanism \cite{Fukugita:1986hr}.
Also, it induces source of lepton flavor violation (LFV). The quantitative analysis of both the phenomena 
(Leptogenesis and LFV) depend on $h_{\nu}$ where $h_{\nu}$ is the Yukawa coupling
of right-handed neutrinos with the SM lepton doublets and Higgs. It is observed that
$h_{\nu}$ must be complex in order to contribute positively to the mechanism like leptogenesis.
To be specific the CP asymmetry parameter, $\epsilon$ depends on imaginary part of $(h_{\nu}h_{\nu}^\dagger)^2$.
$h_{\nu}^\dagger$ appears in the hermitian conjugate part of the extended SM Lagrangian by demanding
that Lagrangians are in general real. Here we study one scenario which allows us
to relax the condition of hermiticity of the Yukawa interaction terms of the Lagrangian while keeping
all the experimental findings related to neutrinos intact.

\begin{table}
 \centering 
 \begin{tabular}{ |p{3cm}||p{3cm}|p{3cm}||p{3cm}|p{3cm}|  }
 \hline
 \multicolumn{5}{|c|}{NuFIT 5.0 (2020)\cite{Esteban:2020cvm}} \\
 \hline
 Neutrino Oscillation parameters& Normal ordering best fit $\pm$ 1$\sigma$ &Normal ordering 3$\sigma$&Inverted ordering best fit $\pm$ 1$\sigma$&Inverted ordering 3$\sigma$\\
 \hline
 \hline
 $\theta_{12}/^{\circ}$   & $33.44^{+0.77}_{-0.74}$    &$31.27\rightarrow35.86$&   $33.45^{+0.78}_{-0.75}$&$31.27\rightarrow35.87$\\
 $sin^{2}\theta_{12}$   & $0.304^{+0.012}_{-0.012}$    &$0.269\rightarrow0.343$&   $0.304^{+0.013}_{-0.012}$&$0.269\rightarrow0.343$\\
 $\theta_{23}/^{\circ}$   & $49.2^{+0.9}_{-1.2}$    &$40.1\rightarrow51.7$&   $49.3^{+0.9}_{-1.1}$&$40.3\rightarrow51.8$\\
 $sin^{2}\theta_{23}$&   $0.573^{+0.016}_{-0.020}$  & $0.415\rightarrow0.616$   &$0.575^{+0.016}_{-0.019}$&$0.419\rightarrow0.617$\\
 $\theta_{13}/^{\circ}$   & $8.57^{+0.12}_{-0.12}$    &$8.20\rightarrow8.93$&   $8.60^{+0.12}_{-0.12}$&$8.24\rightarrow8.96$\\
 $sin^{2}\theta_{13}$ &$0.02219^{+0.00062}_{-0.00063}$ & $0.02032\rightarrow0.02410$&  $0.02238^{+0.00063}_{-0.00062}$&$0.02052\rightarrow0.02428$\\
 $\delta/^{\circ}$    &$197^{+27}_{-24}$ & $120\rightarrow369$&  $282^{+26}_{-30}$&$193\rightarrow352$\\
 $\frac{\Delta m^{2}_{21}}{10^{-5}eV^{2}}$&   $7.42^{+0.21}_{-0.20}$  & $6.82\rightarrow8.04$&$7.42^{+0.21}_{-0.20}$&$6.82\rightarrow8.04$\\
 $\frac{\Delta m^{2}_{3l}}{10^{-3}eV^{2}}$& $+2.517^{+0.026}_{-0.028}$  & $+2.435\rightarrow+2.598$   &$-2.498^{+0.028}_{-0.028}$&$-2.581\rightarrow-2.414$\\
 \hline
\end{tabular}
\caption{Allowed ranges of neutrino oscillation parameters for normal and inverted mass ordering}
\label{neop}
\end{table}

The Lagrangian of the SM is constructed based on three principles;
locality, Lorentz invariance in vacuum and hermiticity. The hermiticity of the
Lagrangian/Hamiltonian ensures real energy eigenvalues and unitarity of time
evolution. In standard texts of Quantum Mechanics one often deals with Hamiltonians
which are hermitian because of the preceding reason. 
It has been shown in \cite{Bender:1998ke}, \cite{Bender:1998gh} ,\cite{Bender:2005tb} that hermiticity is a mathematical requirement
to achieve real eigenvalues from the Hamiltonian. Further, a rather non-Hermitian Hamiltonian
which is symmetric under the combined operation of space reflection operator, ${\mathcal{P}}$
and the time inversion operator ${\mathcal{T}}$ has real eigenvalues provided ${\mathcal{PT}}$
symmetry is unbroken\footnote{Under ${\mathcal{P}}$ the two fundamental operators, $\hat{x}$ and $\hat{p}$ transform like 
$\hat{x}\longrightarrow - \hat{x}$ and $\hat{p}\longrightarrow - \hat{p}$. The time inversion operator has the
effect $\hat{x}\longrightarrow - \hat{x}$, $\hat{p}\longrightarrow - \hat{p}$ and $i \longrightarrow -i$ \cite{Bender:2005tb} . }. 
Thus ${\mathcal{PT}}$ symmetry is an alternative requirement to hermiticity
for constructing many new Hamiltonians which would have been rejected in the past on the basis of
hermiticity.  For example consider a $(2\times 2)$ complex Hamiltonian which is not hermitian but ${\mathcal{PT}}$ symmetric,
\begin{equation}
 \begin{pmatrix}
r e^{i \theta} & \sigma\\
\sigma   & r e^{-i \theta}
\end{pmatrix},
\end{equation}
where the $r, \sigma$ and $\theta$ are real parameters. The eigenvalues are given by 
$r \cos \theta \pm \sqrt{\sigma^2 - r^2 \sin^2 \theta }$. The
eigenvalues are real in one of the parametric region $\sigma^2 > r^2 \sin^2 \theta$ and are
complex for $\sigma^2 < r^2 \sin^2 \theta$.  A construction of $(3\times 3)$ ${\mathcal{PT}}$ symmetric
Hamiltonian is given in \cite{Deng:2012yw},\cite{Wang:2010br}.

Taking a step ahead of \cite{Bender:1998ke},\cite{Bender:1998gh},\cite{Bender:2005tb} the non-Hermitian version of 
Quantum Field Theory has been studied in \cite{Bender:2004vn},\cite{Bender:2004sa}.
Also, a non-Hermitian version of a Quantum Electrodynamics (QED) has been studied in \cite{Bender:2005zz},\cite{Bender:2005hf}.
The authors of \cite{Bender:2005zz}, \cite{Bender:2005hf} consider a minimal non-Hermitian extension of QED in 
the modified Lagrangian,
\begin{equation}
 {\mathcal{L_0}} \supset \bar{\psi}(i \slashed{\partial} - m -\mu \gamma_5) \psi.
 \label{eq:nh-Lagrangian}
\end{equation}
The variant of the Lagrangian has the Hermitian term, $m\bar{\psi}\psi$
along with anti-Hermitian mass term, $\mu \bar{\psi}\gamma_5 \psi$. 
The later term changes sign on Hermitian conjugation. Using the usual
definition of charge conjugation $({\mathcal{C}})$, parity $({\mathcal{P}})$ and time reversal $({\mathcal{T}})$,
it is observed that the anti-Hermitian term is ${\mathcal{C}}$ even, ${\mathcal{P}}$ odd and ${\mathcal{T}}$ even.
So the Lagrangian is ${\mathcal{CP}}$ odd and ${\mathcal{CPT}}$ odd. Nevertheless by varying the above
Lagrangian with respect to $\bar{\psi}$, one can obtain the dispersion relation
as $\omega^2 = p^2 + M^2; M^2 = m^2-\mu^2$. The energies are real when $m^2 \ge \mu^2$.
Also the dispersion relation is real for both particles and antiparticles, thereby
implying no ${\mathcal{CPT}}$ violation. Note that the Lagrangian has a non-hermitian mass term but
the definition used for ${\mathcal{C,P}}\, {\rm and} \,{\mathcal{T}}$ transformations are relevant to a Hermitian
theory. Hence it is appropriate to construct alternate definition of these transformations
compatible with non-Hermitian theory. The appropriate construction of ${\mathcal{C}}$ operator is studied
in \cite{Bender:2005zz} and is shown in \cite{Bender:2005hf} that under the modified definition
of ${\mathcal{C,P}}\, {\rm and} \,{\mathcal{T}}$ transformation, the theory is ${\mathcal{C}}$
even, ${\mathcal{P}}$ odd and ${\mathcal{T}}$ odd. Hence the theory is ${\mathcal{CP}}$ odd, ${\mathcal{PT}}$ and ${\mathcal{CPT}}$ even and thus
falls under the category of ${\mathcal{PT}}$ symmetric Quantum Field Theories. Also in the limit $\mu \rightarrow m $, the fermion is effectively massless. In terms of $\psi_R = \frac{1}{2}(1+\gamma^5) \psi$ and $\psi_L = \frac{1}{2}(1-\gamma^5) \psi$, the Lagrangian can be written as
\begin{equation}
 \mathcal{L_{\rm ferm}} = \bar{\psi_R} i \slashed{\partial}\psi_R + \bar{\psi_L} i \slashed{\partial}\psi_L - \bar{\psi_L}( m +\mu)\psi_R -  \bar{\psi_R}( m -\mu)\psi_L.
\end{equation}
The equations of motion in the limit $ \mu =\pm m$ are
\begin{eqnarray}
\nonumber
 i\slashed{\partial} \psi_L = 2 m \psi_R, && \bar{\psi_R} i\slashed{\partial} \psi_R =0, \, (\mu = m),\\
  i\slashed{\partial} \psi_R = 2 m \psi_L, && \bar{\psi_L} i\slashed{\partial} \psi_L =0, \, (\mu = -m).
\end{eqnarray}
So in the specific limits of $\mu$ one loses half of the degrees of freedom. These points correspond to the boundaries in parameter space beyond which 
${\mathcal{PT}}$ symmetry is broken.

Hermiticity is not a necessary constraint on the up- and/or down- type mass matrix
for any theory  of fermion mass generation \cite{Fritzsch:1999ee}. 
It has been shown in general four zero texture of lepton and quark Yukawa  matrix 
without the assumption of hermiticity in \cite{Barranco:2012ci} is consistent with
results related to neutrino oscillation. The possibility of having the non-hermitian 
Yukawa coupling was studied in \cite{BRANCO1994390} and it is shown that a complete classification
Yukawa coupling with texture zeros has to include non-hermitian structure as well.
A non-Hermitian Yukawa model with an aim to explore the implications in neutrino
sector of the SM has been studied in \cite{Alexandre:2015kra}. Similarly, \cite{Korchin:2016rsf}
has given a model for non-Hermitian Yukawa interaction for
one fermion generation. More particularly they study the effects of the Yukawa interaction
in the Higgs boson decay to a pair of $\tau-$leptons. Also adding
non-Hermitian ${\mathcal{PT}}$-symmetric Hamiltonian and taking matter effect into account,
two flavor neutrino oscillation has been studied in \cite{Ohlsson:2015xsa}. It has been shown
that such consideration can give rise to sub-leading effects in neutrino oscillation
provided the additional parameters are small. The references \cite{Mavromatos:2020bbq},\cite{Alexandre:2020bet},\cite{Alexandre:2020tba},\cite{Mavromatos:2020hfy} study PT symmetric consistent Quantum Field theories containing anti-hermitian Yukawa interactions between axions and right handed neutrinos. The antihermitian Yukawa coupling are assumed to be generated from non-perturbative
effect on the context of string -inspired low energy effective field theories.

In this work, we study a special case of Type-I seesaw mechanism where the 
the Yukawa interaction of the neutral leptons comprise of a hermitian part and a non-hermitian part as inspired by a Lagrangian like given in eq.(\ref{eq:nh-Lagrangian}). We calculate the small neutrino masses for three generation case and extend the analysis to gauge the implications for other phenomenological aspects related to neutrinos. While allowing for generation of Dirac mass term through Higgs mechanism, the couplings are
constrained from experimental data. Also, the analysis shows avenue for low scale seesaw.

In section \ref{sec:qed} we recapitulate the gauge symmetry of non-hermitian extension of extended SM Lagrangian given in eq.(\ref{eq:nh-Lagrangian})\cite{Alexandre:2015kra}. The outcomes are generalized for neutrino masses in the context of Type-I seesaw mechanism in section \ref{sec:nonhermYukawa}. We calculate the three mass eigenvalues for three generation case and 
constraints form results arising from oscillation experiments. In section \ref{sec:fourzero} we study the four zero texture of nonhermitian Yukawa coupling. We diagonalize the nonhermitian mass matrix and extract
the neutrino oscillation parameters.
In section \ref{sec:disc} we discuss our results. Since the non-hermitian parameters will introduce new sources of $CP$ violation in the lepton sector, we discuss constraints
on them in sections \ref{sec:cpv} and \ref{sec:lgnsis}. In section \ref{sec:concl} we conclude with future scope of extending the present work.

\section{Gauge symmetry of nonhermitian extension of Quantum Electrodynamics }
\label{sec:qed}
In this section we summarize the important points of gauge symmetry of the nonhermitian extension of QED as developed in 
\cite{Bender:2005hf} \cite{Alexandre:2015oha} \cite{Alexandre:2015kra}
and the implication in connection with light neutrino masses. In this case the variant of the lagrangian for a four component Dirac spinor as compared to eq.(\ref{eq:nh-Lagrangian}) is given by,
\begin{equation}
 \mathcal{L} = \bar{\psi} \left( i \slashed{\partial} - \slashed{A}
 ( g_V + g_A \gamma^5) - m - \mu \gamma^5 \right) \psi - \frac{1}{4}F^{\mu\nu} F_{\mu\nu},
\end{equation}
where $ F_{\mu\nu} = \partial_\mu A_\nu - \partial_\nu A_\mu$. Here the couplings of the $U(1)$ gauge field includes both vector and axial vector. Whereas the gauge invariance is lost in the massive case $m\neq 0$ and/or $\mu \neq 0$, in the massless limit, $m = \mu =0$, the Lagrangian is invariant under the combined vector and axial vector gauge transformation
 \begin{eqnarray}
  A_\mu  && \rightarrow A_\mu - \partial_\mu \phi,\\
  \psi && \rightarrow {\rm exp}(i (g_V + g_A \gamma^5) \phi) \psi,\\
  \bar{\psi} && \rightarrow {\rm exp}(i(-g_V + g_A \gamma^5)\phi).
 \end{eqnarray}
By varying the Lagrangian with respect to $\bar{\psi}$ the Dirac equation is given by,
\begin{equation}
 (i\gamma^\mu D_\mu - m -\mu \gamma^5)\psi = 0.
\end{equation}
The dispersion relation can be obtained by acting again the Dirac operator,
\begin{equation}
 (D^2 -m^2 -\mu^2)\psi = 0
\end{equation}
and the energy is given by
\begin{equation}
 \omega^2 = p^2 + M^2, \quad M^2 = m^2 -\mu^2.
\end{equation}
The energies are real for $m^2 \ge \mu^2$. Also the equation ensures
$\mathcal{CPT}$ invariance as the dispersion relation is same for particles and antiparticles.

 The free fermion propagator of this theory is given by
\begin{equation}
iS = i \frac{\slashed{p}+m -\mu \gamma^5}{p^2 -M^2 +i\epsilon}.   
\end{equation}
Immediately one can observe that there is a light-like pole for $\mu =\pm m$  and the form of the propagator is given by,
\begin{equation}
 iS = i \frac{\slashed{p} +m(\mathcal{I}\mp \gamma^5)}{p^2 + i \epsilon}.
\end{equation}
The mass term in the propagator is proportional to the chiral projection operator $(\mathcal{I} \pm\gamma^5)/2$. Hence one can work in a specific chiral basis and the Lagrangian can be written as
 \begin{equation}
  \mathcal{L}_{mass} = 
  \begin{pmatrix}
   \psi_L^{\dagger} & \psi_R^{\dagger}
  \end{pmatrix}
  \begin{pmatrix}
   i \bar{\sigma}.D_{-} & -m_+\\
-m_{-}   & i \sigma.D_+
  \end{pmatrix}
\begin{pmatrix}
  \psi_L^{\dagger}\\
   \psi_R^{\dagger}
  \end{pmatrix},
 \end{equation}
where $m_{\pm} = m\pm \mu$, and $D_{\pm} = \partial^\mu + (g_V \pm g_A) A^\mu$. The probability density is given as
\begin{equation}
 \rho = \left( 1+ \frac{\mu}{m}\right) |\psi_R|^2
 + \left( 1 - \frac{\mu}{m} \right) |\psi_L|^2.
\end{equation}
Also the mass matrix of the fermion has the form
\begin{equation}
 {\bf m} = \begin{pmatrix}
            0 & m_+\\
            m_-& 0
           \end{pmatrix}.
\end{equation}
From this analysis we can observe the chirality flip in the following manner. For $\mu = +m (-m)$ we obtain massless theory with chirality flips from left(right) to right(left). Also for $\mu = +m $ the conserved current $\rho = 2|\psi_R|^2$ and the left-chiral current decouples. Similarly,  $\rho = 2|\psi_L|^2$ for  $\mu = -m$ and the right current decouples. This can be understood in an alternate manner. Consider the Lagrangian for $\mu = +m$,
\begin{equation}
 \mathcal{L} = \psi_L^{\dagger} i \bar{\sigma}.D_- \psi_L
 + \psi_R^{\dagger} i \sigma.D_+ \psi_R - 2m \psi_L^{\dagger} \psi_R.
\end{equation}
The Weyl equation is given as
\begin{eqnarray}
 i \sigma.D_+ \psi_R =0, \quad
 i \bar{\sigma}.D_- \psi_L = 2m \psi_R.
\end{eqnarray}
 The left chirality can be integrated out to give on-shell Lagrangian for a massless right chiral Weyl fermion as
\begin{equation}
 \mathcal{L_{\rm on-shell}} = \psi_R^{\dagger} i \sigma.D_+\psi_R.
\end{equation}
In this limit the vector plus axial vector gauge invariance is recovered for 
\begin{equation}
 A_\mu \rightarrow A_\mu - \partial_\mu \phi, \quad
 \psi_R \rightarrow {\rm exp}(i (g_v+g_A) \phi) \psi_R.
\end{equation}
Similarly for $\mu=-m$ one can get massless left chiral theory by integrating out right chiral field invariant under gauge transformation
\begin{equation}
 A_\mu \rightarrow A_\mu - \partial_\mu \phi, \quad
 \psi_L \rightarrow {\rm exp}(i (g_v-g_A) \phi) \psi_R.
\end{equation}
This bearing persists at all loop orders as shown in \cite{Alexandre:2015kra} and can be summarized as, the right chirality dominates for
$0 < \mu \le +m$ and the left chirality dominates for $-m \le \mu <0$.
It is appealing to consider the generalisation of the nonhermitian behavior for light neutrinos in the Higgs-Yukawa theory.
\section{Neutrino mass from non-Hermitian extension of neutral fermionic Lagrangian}
\label{sec:nonhermYukawa}
In this section, we summarize the important results 
related to neutrino masses coming from a non-hermitian Higgs-Yukawa model. And then calculate
the masses for three generation case. Adding right-handed neutrinos, $N_R$'s
one per generation to the SM the Lagrangian for the lepton sector the
nonhermitian neutrino Yukawa sector is given as,
\begin{eqnarray}
 \mathcal{L} &=& \bar{l}_{L,k} i \slashed{D} l_{L,k}\bar{N}^{\alpha}_{R} i \slashed{\partial}
 N_{R,\alpha} - [h_{-}]_k^{\alpha} \bar{l}_L^k \tilde{H} N_{R, \alpha} -
 [h_{+}]^k_{\alpha} \bar{N}_{R}^{\alpha} \tilde{H^{\dagger}} l_{L,k} ,
\label{eq:Dirac-L}
 \end{eqnarray}
%
where $l^k_L \equiv  (\nu^k_L e^k_L)^T$ are the left-handed lepton doublets, $l_R$'s are the
right-handed charged leptons and $\alpha, k$ are generation indices.
The Higgs doublet is represented as $H$ and $\tilde{H} \equiv i \tau_2 H^*$. $D_\mu$
is the covariant derivative of the electroweak gauge group of the SM. The Yukawa couplings $h_{\pm}$
 are given by,
 \begin{equation}
 h_{\pm} = h\pm \eta,  
 \end{equation}
where $h$ and $\eta$ are complex valued matrices. The left and right handed sectors transform under a general transformation in $U_L(N)\times U_R(N)$ as
 \begin{eqnarray}
  l_{L,k} \rightarrow l'_{L,k} = V_k^m l_{L,m} ,& & \quad l^k_L \equiv (l_{L,k})^\dagger \rightarrow {l'_L}^k = V^k_m l_L^m,\\ \nonumber
  N_{R,\alpha} \rightarrow N'_{R,\alpha} = U^\beta_\alpha N_{R,\beta}, &&
  N_R^\alpha \equiv (N_{R,\alpha})^\dagger \rightarrow {N'_{R}}^\alpha  = U^\alpha_\beta N_R^\beta.
 \end{eqnarray}
 where $V_l^m \in U_L(N) $ and $U^\beta_\alpha \in U_R(N) $. The Yukawa matrices transform as 
 \begin{equation}
  [h_\pm]_k^\alpha \longrightarrow [h'_\pm]_k^\alpha = V_k^m U^\alpha_\beta 
  [h_\pm]_m^\beta.
  \label{eq:transform}
 \end{equation}
 
 After spontaneous symmetry breaking due to the
 Higgs field acquiring vacuum expectation value,
 \begin{eqnarray}
  H= \frac{1}{\sqrt{2}}
  \begin{pmatrix}
   0\\
   \upsilon + \xi
  \end{pmatrix}, & & 
  \tilde{H} = \frac{1}{\sqrt{2}}
  \begin{pmatrix}
    \upsilon + \xi\\
    0
  \end{pmatrix},
 \end{eqnarray}
 The Lagrangian can now be written in Dirac basis as,
 \begin{equation}
  \mathcal{L_\nu} \supset 
  \begin{pmatrix}
   \bar{\nu}_L & \bar{N}_L
  \end{pmatrix}
  \begin{pmatrix}
   i \slashed{\partial} & - h_- \frac{\upsilon}{\sqrt{2}}\\
   - h_+ \frac{\upsilon}{\sqrt{2}} & i \slashed{\partial}
  \end{pmatrix}
  \begin{pmatrix}
   \nu_L\\
   N_R
  \end{pmatrix}.
 \end{equation}
The mass matrix of the neutrinos can be written as
\begin{equation}
  M = \frac{\upsilon}{\sqrt{2}}
\begin{pmatrix}
  0 & h_-\\
  h_+ & 0
\end{pmatrix}.
\end{equation}
The mass eigenvalues for two generation
 case $(N =2)$ are given as the roots of
 \begin{equation}
  m^2_{1(2)} = \frac{\upsilon^2}{4}\left[ {\rm tr}~h_{+}^\dagger h_{-} 
 -(+) \left( 2 {\rm tr}~\left(h_{+}^\dagger h_{-}\right)^2-
 ({\rm tr}~h_{+}^\dagger h_{-})^2\right)^{1/2}\right].
 \end{equation}
There are two conditions to get a massless spectrum for the neutrinos;
\begin{enumerate}
 \item by choosing $h = \pm \eta$,
 \item in the case ${\rm det}~(h_{+}^\dagger h_{-}) = 0$.
\end{enumerate}
For the later case the we get two massless $(m_1^2 =0)$ states and two massive states,
\begin{equation}
m_2^2 = \frac{\upsilon^2}{2}  {\rm tr}~h_{+}^\dagger h_{-} =
\frac{\upsilon^2}{2} \left[ {\rm tr}~h^\dagger h - {\rm tr}~\eta^\dagger \eta - 2 i ~{\rm Im\, tr}~ h^\dagger \eta\right].                                                                                    \end{equation}
Further $m_2$ will be real for vanishing trace of $(h^\dagger \eta)$ and one can 
get vanishing $m_2$ for,
\begin{equation}
 {\rm tr ~h^\dagger h} = {\rm tr ~\eta^\dagger \eta}.
 \label{eq:}
\end{equation}
Also one can get finite and sub-eV Dirac neutrino mass by choosing
\begin{equation}
 {\rm tr}~ h^\dagger h \sim {\rm tr}~ \eta^\dagger \eta.
\end{equation}
The complex matrices $h$ and $\eta$ have total of $16$ parameters (for $2$ generation case) but relations obtained above impose weaker constraints on the
elements than the condition $h=\eta$. In the coming sections we explore
the possibility of constraining the elements of $h$ and $\eta$ using neutrino oscillation parameters.

\subsection{Neutrino mass generation through type-I seesaw mechanism}
The right handed neutrinos are singlets of the SM gauge groups so,
a Majorana mass term is inevitable in the Lagrangian.  The new term can be added to the Lagrangian given in eq.(\ref{eq:Dirac-L}) can be written as, 
\begin{equation}
 -\left( \frac{1}{2} \bar{N}^C_{R,\alpha} M_R^{\alpha \beta} \bar{N}_{R \beta}+ {\rm h.c.}\right),
\end{equation}
where $C$ denotes the charge conjugation.
In the basis $(\nu^C_L \,\, N_R)^T$ now the Lagrangian ca be written as,
\begin{eqnarray}
\nonumber
 -\mathcal{L_{\nu}} &\supset& \frac{1}{2} (\bar{\nu}^k_L  ~~~~ \bar{N}^C_{R,\alpha})
  \begin{pmatrix}
  0  & [m_{-}]^{\beta}_k\\
  [m_{+}]^{\alpha}_l & M_R^{\alpha\beta}
 \end{pmatrix}
 \begin{pmatrix}
  \nu_L^{C,L}\\
  N_{R,\beta}
 \end{pmatrix} \\
 &+& \frac{1}{2}  (\bar{\nu}^C_{L,k}  ~~~~ \bar{N}^{\alpha}_{R})
  \begin{pmatrix}
  0  & [m_{-}]_{\beta}^k\\
  [m_{+}]_{\alpha}^l & M_{R, \alpha\beta}
 \end{pmatrix}
\begin{pmatrix}
  \nu_{L,l}\\
  N_{R}^{C,\beta}
 \end{pmatrix},
 \label{eq:langfull}
\end{eqnarray}
where 
\begin{equation}
 m_{\pm} = m \pm \mu =\frac{\upsilon}{\sqrt{2}} (h\pm \eta),
 \label{eq:mpm}
\end{equation}
where $m = \upsilon h/\sqrt{2}$ is the hermitian mass term and $\mu = \upsilon \eta /\sqrt{2}$
is the antihermitian mass term. From the above equation the mass matrix of the light and heavy neutrinos can be written as,
\begin{equation}
 M = 
 \begin{pmatrix}
  0 & m_{-}\\
  m_{+}^T & M_R
  \label{eq:seesawMass}
 \end{pmatrix}.
\end{equation}
Further one can block diagonalize the mass matrix by a unitary transformation,
 $M^{\rm diagonal} = W^T M W$.
This would yield the mass matrix of the light neutrinos and is given by,
\begin{equation}
 m_L = -\frac{\upsilon^2}{2} (h_{+}^T M_R^{-1} h_{-}).
 \label{eq:mL}
\end{equation}
It is interesting to note that this seesaw formula is a non-Hermitian one in 
contrast to the usual cases. From eq.(\ref{eq:mL}) the individual mass eigenvalues can be found 
for physical neutrinos.
\subsection{Case for two generation of neutrinos}
For two generation $h$ and $\eta$ are $(2\times2)$ complex matrices. Hence there are total $16$
parameters which can be constrained from experimental data.
The individual neutrino mass eigenvalues for $N=2$ is given by \cite{powell2011calculating}
\begin{equation}
 m_{1(2)} =-\frac{\upsilon^2}{4}\left[ {\rm tr}h_{+}^T M_R^{-1}h_{-} 
 -(+) \left( 2 {\rm tr}\left(h_{+}^T M_R^{-1} h_{-}\right)^2-
 ({\rm tr}h_{+}^T M_R^{-1}h_{-})^2\right)^{1/2}\right].
 \label{eq:m1m2}
\end{equation}
The non-Hermitian theory would account for a mass less spectrum if
\begin{equation}
h= \pm\eta.
\label{eq:heta1}
\end{equation}
But one can obtain a non-zero mass for neutrinos when,
\begin{equation}
 {\rm det }~ h_{+}^T M_R^{-1}h_{-} = 0 \quad \implies \quad 
 {\rm tr}~\left(h_{+}^T M_R^{-1} h_{-}\right)^2 =
 ({\rm tr}~h_{+}^T M_R^{-1}h_{-})^2.
\end{equation}
With this condition we obtain the spectrum
\begin{equation}
 m_1 = 0, \quad\quad m_2 = -\frac{\upsilon^2}{2}{\rm tr}h_{+}^T M_R^{-1}h_{-} .
\end{equation}
The physical neutrino mass are real i.e. for $m_2$ to be real we require 
\begin{equation}
 {\rm Im~tr }~h_{+}^T M_R^{-1}h_{-} = 0;\quad {\rm Re \, tr}(h_{+}^T M_R^{-1}h_{-}) \neq 0,
\label{eq:heta2}
 \end{equation}
and we can restore a mass less spectrum again if the following condition is met.
\begin{equation}
  {\rm Re~tr }~h_{+}^T M_R^{-1}h_{-} = 0.
  \label{eq:heta3}
\end{equation}
For two generation of neutrinos there are total $16$ parameters
in the complex valued $2\times 2$ Yukawa matrices. The constraints
provided by eq.s.(\ref{eq:heta2}) and (\ref{eq:heta3}) are much
weaker than that given in eq.(\ref{eq:heta1}) in determining the
Yukawa matrices. However neutrino oscillation data will be useful
in provided observational constraints on $h$ and $\eta$.
\subsection{Case for three generation of neutrinos}
\label{sec:3gen}

In order to verify the results of experiments related to neutrino phenomenology,
it is necessary to calculate neutrino masses for three generations of
neutrinos. In this case $h$ and $\eta$ are $(3\times 3)$ complex matrices having 
$18$ real parameters each. The mass eigenvalues of eq.(\ref{eq:mL}) for $N=3$ are given 
by the roots of the equation \cite{powell2011calculating},
\begin{eqnarray}
&& \left( {\rm tr}( m_{+}^T M_R^{-1} m_{-} +\lambda I) \right)^3 - 
 3 {\rm tr}(m_{+}^T M_R^{-1} m_{-} +\lambda I ) \\\nonumber
  && {\rm tr}\left( m_{+}^T M_R^{-1} m_{-} +\lambda I \right)^2
+2  {\rm tr}\left(m_{+}^T M_R^{-1} m_{-} +\lambda I \right)^3 = 0,
\end{eqnarray}
where $m_{\pm}$ is given in eq.(\ref{eq:mpm}). After simplification the
mass eigenvalues are given by,
\begin{eqnarray}
\nonumber
 m_1 &=& - \frac{a}{3} - \frac{2^{1/3}(a^2/2 - 3b/2)}{3 f}+ \frac{f}{3\times 2^{1/3}}, \\ \nonumber
 m_2 &=& - \frac{a}{3} + \frac{(1+ i \sqrt{3})(a^2/2 - 3b/2)}{3 \times 2^{2/3} f} -
 \frac{(1 - i \sqrt{3}) f }{6\times2^{1/3}} ,\\
 m_3 &=& - \frac{a}{3} + \frac{(1 - i \sqrt{3})(a^2/2 - 3b/2)}{3\times2^{2/3} f}
- \frac{(1+ i \sqrt{3}) f }{6 \times 2^{1/3}},
\label{eq:m1m2m3}
 \end{eqnarray}
where 
\begin{eqnarray}
\nonumber
 f &=& \left[ -2 a^3 + 9a b \right.\\ \nonumber 
 &+& \left. \sqrt{4(a^2/2 - 3b/2)^3 + 
 (-2 a^3 + 9a b - 9c)^2} - 9c \right]^{1/3},\\ \nonumber
 a &=&  {\rm tr} \left(m_{+}^T M_R^{-1} m_{-}\right),\\ \nonumber
 b &=& {\rm tr}\left(m_{+}^T M_R^{-1} m_{-}\right)^2,\\ 
 c &=&  {\rm tr}\left(m_{+}^T M_R^{-1} m_{-}\right)^3.
 \label{eq:abcf}
\end{eqnarray}
From eq.(\ref{eq:m1m2m3}) the sum of three neutrino masses can be verified to be,
\begin{equation}
 \sum_i m_i = - a = - {\rm tr}\left(m_{+}^T M_R^{-1} m_{-} \right),
\end{equation}
which can be compared with case $N=2$ given in eq.(\ref{eq:m1m2}). Also 
we can have a mass less spectrum of light neutrinos for $h = \pm \eta$ as discussed in 
the previous subsection.

In eq.(\ref{eq:m1m2m3}) the first term is equal for all three mass eigenvalues,
-$\frac{1}{3}{\rm tr}m_{+}^T M_R^{-1} m_{-} $.  The mass eigenvalues arising from
the non-Hermitian theory can account for both degenerate and hierarchical spectrum
by adjusting the parameters in $h$ and $\eta$. 
The masses are degenerate, $m_1= m_2 = m_3$ for 
\begin{equation}
 f= 2^{1/3}\sqrt{a^2/2 - 3 b/2},
 \label{eq:deg_cond}
\end{equation}
$f$ can not be zero, else masses will be undetermined. In order to comply with
neutrino oscillation data the tiny mass squared difference
can be calculated in the leading order to be,
\begin{eqnarray}
 && \Delta m_{ij}^2 \sim \mathcal{O} ({\rm tr}~m_{+}^T M_R^{-1} m_{-})^2 
 \sim \mathcal{O}\left( \frac{\upsilon^4}{4} ({\rm tr}~h_{+}^T M_R^{-1} h_{-})^2 \right) \\ \nonumber
 && \sim \mathcal{O}( \frac{\upsilon^4}{4}\left({\rm tr }~\left(h^T M_R^{-1}h - \eta^T M_R^{-1} \eta +  
 \eta^T M_R^{-1} h - h^T M_R^{-1} \eta\right) \right)^2).
 \end{eqnarray}
 The neutrino oscillation data can be used to constrain the parameters of $h$ and $\eta$. For example using
 the order of magnitude of atmospheric neutrino mass squared difference $\sim \mathcal{O}(10^{-3})~ {\rm eV}^2$ and using
 $\upsilon \simeq 246~$ GeV we get,
 \begin{eqnarray}
 \nonumber
 && \mathcal{O}\left({\rm tr }~\left(h^T M_R^{-1}h - \eta^T M_R^{-1} \eta +  
 \eta^T M_R^{-1} h - h^T M_R^{-1} 
 \eta\right) \right)^2\\ 
 && \sim 10^{-3} \times \frac{4}{\upsilon^4} {\rm eV}^{-2}\sim \mathcal{O}(10^{-24}) {\rm eV}^{-2}.
   \label{eq:heta24}
 \end{eqnarray}
 
 This indicates that the parameters of $h$ and $\eta$ must accommodate mutual cancellation among
 themselves in order to be consistent with above condition. So we explore the parameter space available for the  parameters of $h$ and $\eta$ taking both of them to be order unity. It is interesting to note that  the parameters are independent of the choice of the scale of $M_R$ as identified by \cite{Alexandre:2015kra}. This opens up a new possibility for realization of seesaw mechanism at low energy scale. But for three generation of leptons the number of complex parameters are way larger than the number of constraints. So it will be useful to impose some symmetry to explore the parameter space. In the next section we undertake the phenomenological study using four zero texture of leptonic Yukawa couplings.

\section{Four zero texture of non hermitian Yukawa couplings}
\label{sec:fourzero}

The transformation in eq.(\ref{eq:transform}) leaves the nonhermitian Yukawa coupling in an arbitrary complex form,
\begin{equation}
 h_{\pm ij} = c_{\pm ij} e^{i\, \phi_{\pm ij}}.
\end{equation}
The nonhermitian nature of Yukawa coupling matrices gives rise to an enhanced number of the unknown parameters of the theory than their hermitian or complex symmetric counterpart. Hence, they throw main challenges in phenomenological study. Further nonhermitian mass matrices have been studied in the context of 
for nearest neighbor mixing \cite{Branco:1988iq},\cite{BRANCO1994390}, in terms of triangular form of mass matrices \cite{Haussling:1997ue}, pure phase matrices based on hypothesis of universal strength of Yukawa matrices \cite{Branco:1996fb},\cite{Branco:1990fj} and predicting CKM matrix and universal seesaw mechanism in the context of universal strength Yukawa coupling in \cite{Branco:1995pw} and \cite{Shinohara:1998bw} respectively. Here we explore the possibility by incorporating texture zeros in the coupling matrices which reduces the number of free parameters thereby making the theory predictive. Four zero texture of leptonic mass matrices have been successful in explaining neutrino oscillation data. The hermitian form of four zero texture has been studied in \cite{Nishiura:1999yt} and the nonhermitian form in \cite{Barranco:2012ci}.

It is possible to enforce zeros in the Yukawa matrix by by imposing abelian  flavor symmetries at the expense of extra fields to the basic field content \cite{Grimus:2004hf}. Also with help of weak basis (WB) transformation it has been shown to get texture zeros in the quark sector \cite{Branco:1999nb} and in the lepton sector \cite{Branco:2007nn}. Here to begin with the Yukawa matrices have arbitrary complex structure and by suitable WB transformation one can obtain set of zeros in the mass matrix. It is further shown in \cite{Branco:2007nn} that starting from arbitrary lepton mass matrices it is possible to obtain some of the set of zeros which has no physical content. But in order to realize four zero texture where the charged lepton (cl) and neutrino mass matrix have the same texture, it is not possible to obtain same set of zeros simultaneously through WB transformation demanding that the charged lepton mass matrix, $m_{\rm cl}$ hermitian. Hence one has to relax the hermiticity condition on $m_{\rm cl}$ and hence implying some physical implication.

Moreover since the neutrino mass originates from seesaw mechanism the neutrino mass matrix also should retain the same texture. The seesaw realization of four zero texture has been implemented  in \cite{Barranco:2010we},\cite{Branco:2007nn},\cite{Nishiura:1999yt}. This can be achieved by imposing same texture zeros in right handed neutrino mass matrix as Dirac mass matrix. Keeping these motivation in mind, the non-hermitian four zero texture is given by,
\begin{equation}
 m_+ = \frac{\upsilon}{\sqrt{2}} h_+ = \frac{\upsilon}{\sqrt{2}} 
 \begin{pmatrix}
  0 & h_1 \, e^{i\, \phi_1} & 0\\
  h_1 \, e^{i\, \phi_2 } & h_2 \, e^{i\, \phi_3 } & h_4 \, e^{i\, \phi_4 }\\
  0 & h_4 \, e^{i\, \phi_5 } & h_3 \, e^{i\, \phi_6 }
 \end{pmatrix},
 \label{eq:four-zero}
 \end{equation}
where $h_+$ is given by,
\begin{equation}
 {h_+}_{ij} = h_{ij} \,+\, \eta_{ij}.
\end{equation}
To be specific we can write the $(3,3)$ element of $h_+$ matrix as
\begin{equation}
 h_3 \, e^{i\, \phi_6} = h_{33} + \, \eta_{33},
\end{equation}
where $h_{33}$ and $\eta_{33}$ are the $(3,3)$ elements of $h$ and $\eta$ respectively and both are complex. Now in order to realize the seesaw origin as well as four zero texture of light neutrino mass matrix we assume the right handed neutrino mass matrix also take four zero texture form. This kind of analysis has been explicitly 
verified in \cite{Nishiura:1999yt,Barranco:2010we}. The right handed neutrino mass matrix is given by
\begin{equation}
M_R=
 \begin{pmatrix}
  0 & C_R & 0\\
  C_R & \tilde{B}_R & B_R\\
  0 & B_R & A_R
 \end{pmatrix},
\end{equation}
where the entries in the above matrix are real and the matrix is symmetric. With these textures the light neutrino mass matrix is given by
\begin{equation}
 m_L = 
 \begin{pmatrix}
  0 & C_\nu & 0\\ 
 C^{\prime}_\nu & D_\nu & B_\nu\\
    0 & B^{\prime}_\nu & A_\nu
 \end{pmatrix}.
 \end{equation}
The elements of the matrix are given as
\begin{eqnarray}
\nonumber
 C_\nu &=& -\frac{\upsilon^2}{2}  \frac{h_1^2}{C_R},\\ \nonumber
  C'_\nu &=& - \frac{\upsilon^2}{2}\frac{h_1^2}{C_R},\\ \nonumber
 D_\nu &=& \frac{\upsilon^2}{2}\frac{1}{A_R\, C_R^2} \left( h_1^2\, A_R \,\tilde{B_R}\, (e^{i\,(\phi_1-\phi_2)} +  \, e^{i\,(\phi_3-\phi_2)}) \right. \\ \nonumber
 &+& \left.h_1 \, h_4 \, ( e^{i\,(\phi_1-\phi_4)} +  \, e^{i\,(\phi_5 -\phi_2)}) - h_4^2\, e^{i\, (\phi_5-\phi_4)} \right ),\\ \nonumber
 B_\nu &=& \frac{\upsilon^2}{2} \frac{1}{A_R\, C_R}\left(  D_R\,h_1\,h_3\, e^{i\,(\phi_6-\phi_2)} - A_R\,
  h_1 \,h_4\, e^{(\phi_4 \,-\,\phi_2)} \right. \\ \nonumber
  &-& \left. C_R\,h_3\,h_4\, e^{(\phi_6\,-\,\phi_4)}
  \right),\\ \nonumber
 B'_\nu &=& \frac{\upsilon^2}{2} \frac{1}{A_R\, C_R}\left(  D_R\,h_1\,h_3\, e^{i\,(\phi_1-\phi_6)} - A_R\,
  h_1 \,h_4\, e^{(\phi_1 \,-\,\phi_5)} \right. \\ \nonumber 
  &-& \left. \,C_R\,h_3\,h_4\, e^{(\phi_5\,-\,\phi_6)}
  \right),\\ 
  A_\nu &=& -\frac{\upsilon^2}{2} \frac{h_3^2}{A_R} .
\end{eqnarray}
 
As stated in the beginning of this section the charged lepton mass matrix can be represented as,
\begin{equation}
m_{cl}=
 \begin{pmatrix}
   0 & C_l & 0\\
   C^{\prime}_l & D_l & B_l\\
    0 & B^{\prime}_l & A_l
    \end{pmatrix}.
 \end{equation}
 Here we choose the $(i,j)$ and $(j,i)$ elements of the matrix having same modulus but different phase such as; $C_f(C_f^{\prime})=c_f e^{i(\phi_{cf}(\phi_{cf'}))}, B_f(B_f^{\prime}) = b_f e^{i(\phi_{bf}(\phi_{bf'}))}, D = d e^{i\phi_{df}}$ and $A=ae^{i\phi_{af}}$ for $f =l, \nu$.
 In case of neutrino sector $\phi_c =\phi_{c'} =0$ and $\phi_a=0$.
 
 In order to check the consistency with neutrino data the 
 mass matrices are diagonalised next. In general any complex matrix is diagonalised using biunitary transformation,
 \begin{equation}
 {\rm Diag}(m_{1f},m_{2f},m_{3f}) =  U_L^\dagger m_f U_R.
 \end{equation}
However the unitary matrices are found by solving the equations
\begin{eqnarray}
 U_L^\dagger m_f m_f^\dagger U_L &=& {\rm Diag}(m^2_{1f},m^2_{2f},m^2_{3f}),\\
  U_R^\dagger m_f^\dagger m_f U_R &=& {\rm Diag}(m^2_{1f},m^2_{2f},m^2_{3f}).
 \end{eqnarray}
Here we choose to find the matrix $U_L$ and hence by finding out $h = m_f m_f^\dagger$ and omitting the index $f$ as 
\begin{equation}
    h = 
    \begin{pmatrix} \left|C\right|^2 & CD^* & CB^{\prime *}\\
    DC^* & \left|B\right|^2 +\left|C\right|^2+\left|D\right|^2 & B^{\prime *}D + A^*B\\B^\prime C^* & B^\prime D^* + AB^* &
    \left|A\right|^2 + \left|B\right|^2 
    \end{pmatrix}.
\end{equation}
The diagonalisation of the matrix $h$ can be carried out by the following
procedure \cite{Barranco:2012ci}. 
The $h$ matrix can be factorised in terms of phase matirces as 
 \begin{equation}
  h = P^\dagger \, \widetilde{h}\, P,\,\,  P=e^{-\frac{i}{2} \Xi}diag(e^{\frac{i}{2}\Xi},
    e^{i(\phi_C - \phi_D)},e^{i(\phi_C + \phi_{B^\prime} + \Xi)}),
\label{eq:factor}
\end{equation} and 
\begin{equation}
    \Xi = arctan\left[\frac{a\; sin(\phi_B+\phi_{B^\prime} - \phi_A - \phi_D)}{d+a\; cos(\phi_B+\phi_{B^\prime} - \phi_A - \phi_D)}\right],
\end{equation}
\\
\begin{equation}
    \widetilde{h}= \begin{pmatrix} c^2 & cd & bc\\
    cd & c^2 + d^2 + b^2 & b\left|d+a\delta\right|\\
    bc & b\left|d+a\delta^*\right| & a^2 + b^2 
    \end{pmatrix},
\label{eq:htilde}
\end{equation}
where $\delta = e^{i (\phi_D - \phi_{B'} -\phi_B + \phi_A)}$.

The $\widetilde{h}$ is a real symmetric matrix, hence can be diagonalised by an orthogonal matrix $O_{ij} = v_i(m_j^2)$, where $v_i(m_j^2)$ are the eigenvectors of $\widetilde{h}$ given in the appendix (\ref{app:elem}).  $m_j^2$ are the eigenvalues of $\widetilde{h}$. The orthogonal property of $O$ allows the following invariants of the system which further reduces the number of free parameters.
\begin{equation}
\begin{split}
    Tr(\widetilde{h}) &= m^2_1 + m^2_1 + m^2_3 ,\\
    Tr^2(\widetilde{h}) - Tr(\widetilde{h}^2) &= 2m^2_1m^2_2 + 
    2m^2_1m^2_3 + 2m^2_2m^2_3 ,\\
    Det(\widetilde{h})&=m^2_1 m^2_2 m^2_3 .
    \label{eq:mass-squares}
\end{split} 
\end{equation}
The diagonalizing matrix $U_L$ is thus given by $U_L = O\,P $ as required
by eq.(\ref{eq:factor}).

 From eq.s (\ref{eq:htilde}) and (\ref{eq:mass-squares}) it is clear that the eigen values $m_i^2, (i=1,2,3)$ are independent of other free parameters for $ \Xi = 0$. With this property the the matrix with only phases reduces to $ P = {\rm diag} (1, e^{i\phi_1 },e^{i\phi_2})$, where $\phi_1 = \phi_C - \phi_D, \, {\rm and}\, \phi_2 = \phi_C + \phi_{B^\prime} $.

 We consider normal ordering of light neutrino masses $m_1< m_2 < m_2$ and and by doing so the number of solutions of eq. (\ref{eq:mass-squares}) is reduced to three and there exist three independent parametrizations of $U_L$. We consider one parametrization
\begin{equation}
  \widetilde{m}_1 \le a^\prime \le \widetilde{m}_3  \quad\quad
  \begin{cases}
   b^\prime=
    \sqrt{\frac{(a^\prime-\widetilde{m}_1)(a^\prime+\widetilde{m}_2)(\widetilde{m}_3-a^\prime)}{a^\prime}}\\
    d^\prime= -a^\prime +\widetilde{m}_1 -\widetilde{m}_2 +\widetilde{m}_3\\
    c' = \sqrt{(\tilde{m_1}\tilde{m_2}\tilde{m_3})/a'};
    \end{cases}
\end{equation} 
where the parameters are given as
\begin{equation}
 a' = a/m_3,\, b'=b/m_3,\, 
 \tilde{m_i} = m_i/m_3, i =1,2,3.
\end{equation}
The parameter $b' > 0$ requires the parameter $a'$ to vary between two masses $\tilde{m}_1$ and $\tilde{m}_3$ which can expressed with help of a free parameter $x$ ranging between $0\, {\rm to}\, 1$ as
\begin{equation}
 a' = \tilde{m}_3 \left ( 1- x \frac{\tilde{m}_3 - \tilde{m}_1}{\tilde{m}_3}\right ).
\end{equation}
The above parametrization implies that the parameters $b',c'$ and $d'$
can be expressed in terms of $a'$ and further $a'$ can be varied by varying $x$.

The flavor mixing matrix thus comes from the mismatch between the diagonalization of mass matrices of charged leptons and left handed light neutrinos as
\begin{equation}
 U_{\rm PMNS} = U_l \,P_{l-\nu}\, U_\nu;\quad P_{l-\nu} = {\rm Diag}\left(1,\, e^{(\phi_{1l} - \phi_{1\nu})},\,  e^{(\phi_{2l} - \phi_{2\nu})}\right).
\end{equation}
To restrict the parameter space we choose the phases $\phi_{1l} - \phi_{1\nu} = \phi_{2l} - \phi_{2\nu} = \phi$. So the $U_{\rm PMNS}$ is now explicit function of $a_l', a',m_3$ and $\phi$. 

The angles can be extracted as
\begin{equation}
 \sin{\theta_{13}} = |U_{13}|;\quad
 \tan{\theta_{12}} = \frac{|U_{12}|}{|U_{11}|};\quad \tan{\theta_{23}} = \frac{|U_{23}|}{|U_{33}|},
 \label{eq:angles}
\end{equation}

Since with the determination of $\theta_{13}$ the Dirac phase $\delta$ can generate $CP$ 
violating effects in neutrino oscillations, the magnitude of $CP$ violation is
determined by the the Jarlskog phase invariant is given by,
\begin{eqnarray}
 J_{\rm cp} &=& {\rm Im} \left( U_{11}\,U_{22}\,U^*_{12} \, U^*_{21}\right) \\ \nonumber 
 &=& \frac{1}{8} \cos\,\theta_{13}\,\sin\,2\theta_{12}\, \sin\, \theta_{23}\,\sin\,2\theta_{13} \sin\,\delta.
 \label{eq:jcp}
\end{eqnarray}

\section{Discussion}
\label{sec:disc}
\begin{figure}[h!]
  \centering
  \begin{subfigure}[b]{0.45\linewidth}
    \includegraphics[width=\linewidth]{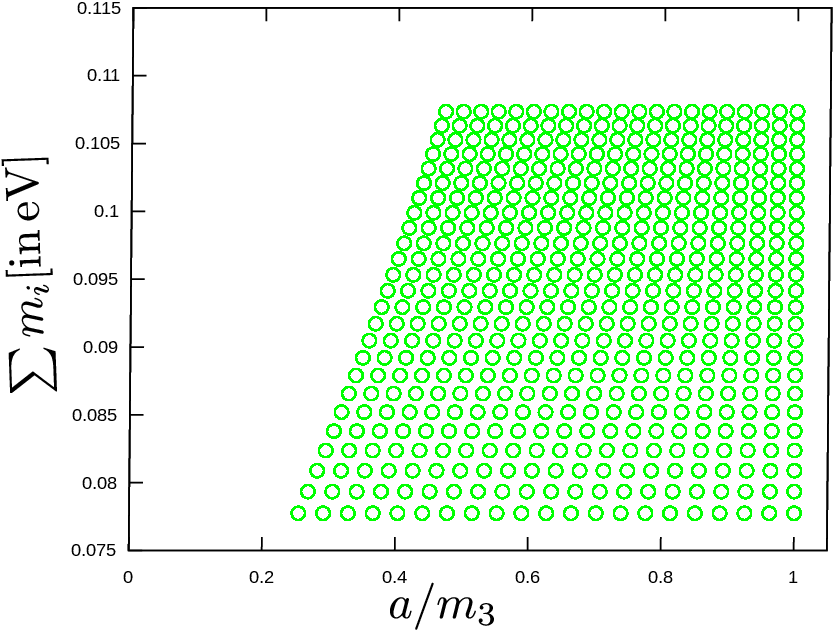}
    \caption{}
     \label{fig:mass-th13-a}
  \end{subfigure}
  \hfill 
  \begin{subfigure}[b]{0.45\linewidth}
    \includegraphics[width=\linewidth]{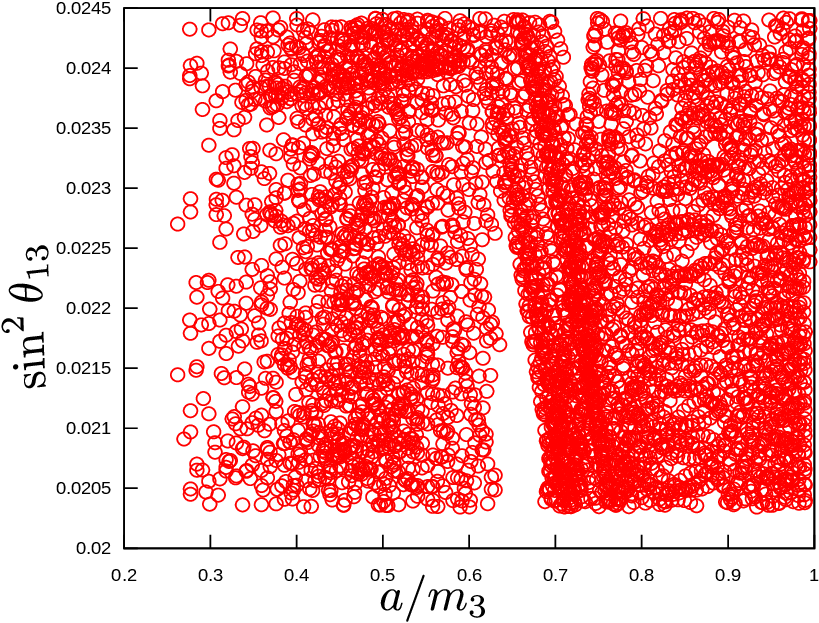}
    \caption{}
     \label{fig:mass-th13-b}
  \end{subfigure}
  \caption{Figure (a) shows the range of sum of three light neutrino masses vs $a'$ which falls
    well within recent Planck limit. Figure (b) shows the values of $\sin^2\theta_{13}$ vs $a'$. }
  \label{fig:mass-th13}
\end{figure}

\begin{figure}[h!]
  \centering
  \begin{subfigure}[b]{0.45\linewidth}
    \includegraphics[width=\linewidth]{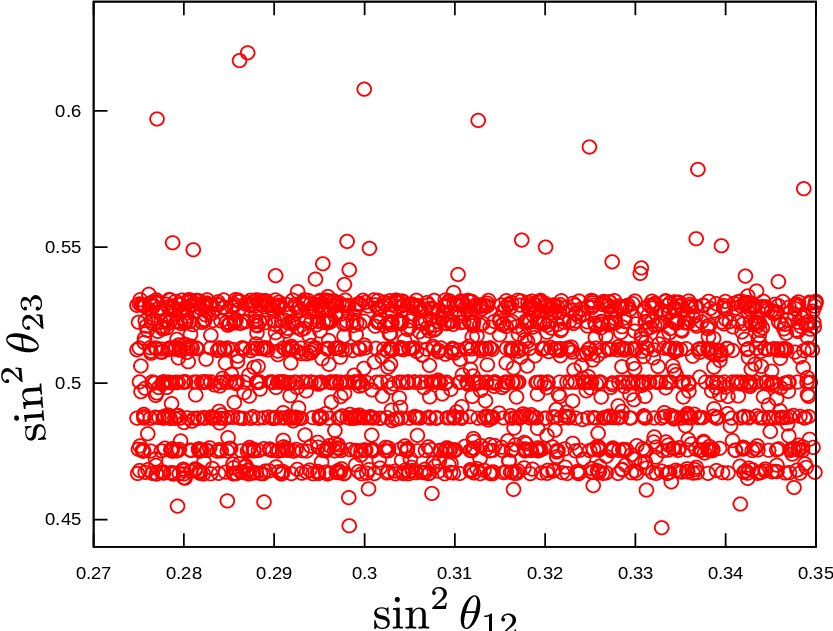}
    \caption{}
    \label{fig:th12-23-a}
  \end{subfigure}
  \hfill
  \begin{subfigure}[b]{0.45\linewidth}
    \includegraphics[width=\linewidth]{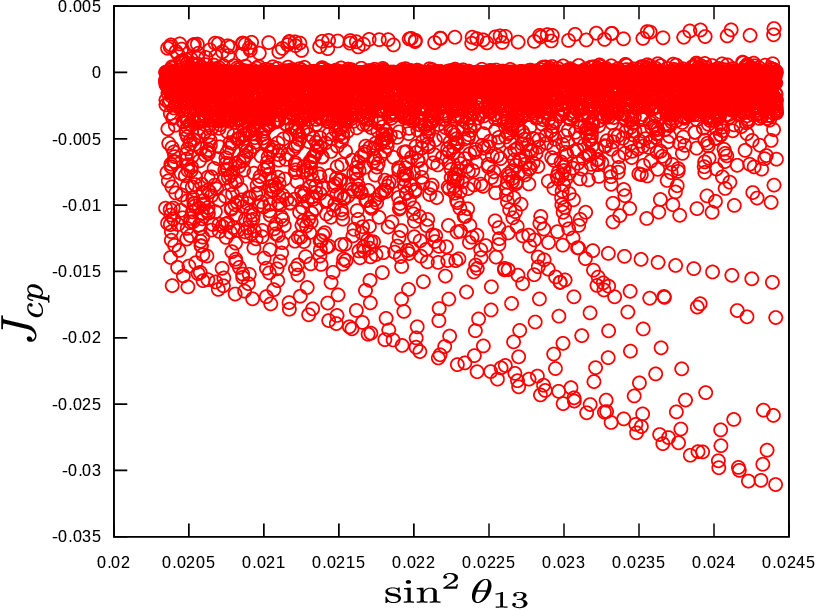}
    \caption{}
    \label{fig:th12-23-b}
  \end{subfigure}
  \caption{In figure (a) the values of $\sin^2\theta_{23}$ and $\sin^2\theta_{12}$ are obtained in the diagonalization process. In figure (b) shows the values of $J_{CP}$ vs. $\sin^2\theta_{13}$. }
  \label{fig:th12-23}
\end{figure}

The flavor mixing matrix, $U_{PMNS}$ depends on $x_f, m_3,\phi$. The charged lepton masses are well determined  and the light neutrino masses are taken from table (\ref{neop}). Taking $m_3$ value around $0.05$ eV the
parameters $x_f$ are varied in the range $0-1$, $\phi$ in the range $0\,-\,2\pi$. The elements of $U_{PMNS}$ matrix is obtained and the three mixing angles and Dirac CP-phase is extracted as given in eq.s (\ref{eq:angles}) and (\ref{eq:jcp}). The values obtained are presented in figures listed as follows. In figure (\ref{fig:mass-th13-a}) the sum of three light neutrino masses meet the recent Planck bound $\sum m_i < 0.12$ eV \cite{Aghanim:2018eyx}. The three mixing angle are also falling in the $3\sigma$ range of latest neutrino global analysis as given in recent data NuFit.org \cite{Esteban:2020cvm} and \cite{deSalas:2020pgw}. The values are shown in figures (\ref{fig:mass-th13-b}) and  (\ref{fig:th12-23-a}). In figure (\ref{fig:th12-23-b}) the $J_{CP}$ values are recorded. Using $3\sigma$ ranges of the three mixing angles the recent data implies $0.030 \sin\,\delta \lesssim J_{CP} \lesssim 0.035 \sin\, \delta$. Using the value of CP violating phase from table (\ref{neop}) the bound can be given by  $-0.0243 < J_{CP} < 0.0037$. 


In our numerical work we find all the mixing angles are generated around the value $a' = a/m_3=0.650$. This  constraint can be used to get the parameter space allowed by $h_{33}$ and $\eta_{33}$ for specific values of $A_R$. As identified in eq.(\ref{eq:heta24}) the mass squared difference is independent of the  right handed neutrino mass scale. The constraints on nonhermitian Yukawa matrix elements comes through 
\begin{equation}
 a' = \frac{a}{m_3}= \frac{h_3^2}{A_R} = \frac{v}{\sqrt{2}} \frac{|h_{33}|^2 + |\eta_{33}|^2}{m_3 A_R}.
\end{equation}

\begin{figure}[h!]
  \centering
  \begin{subfigure}[b]{0.45\linewidth}
    \includegraphics[width=\linewidth]{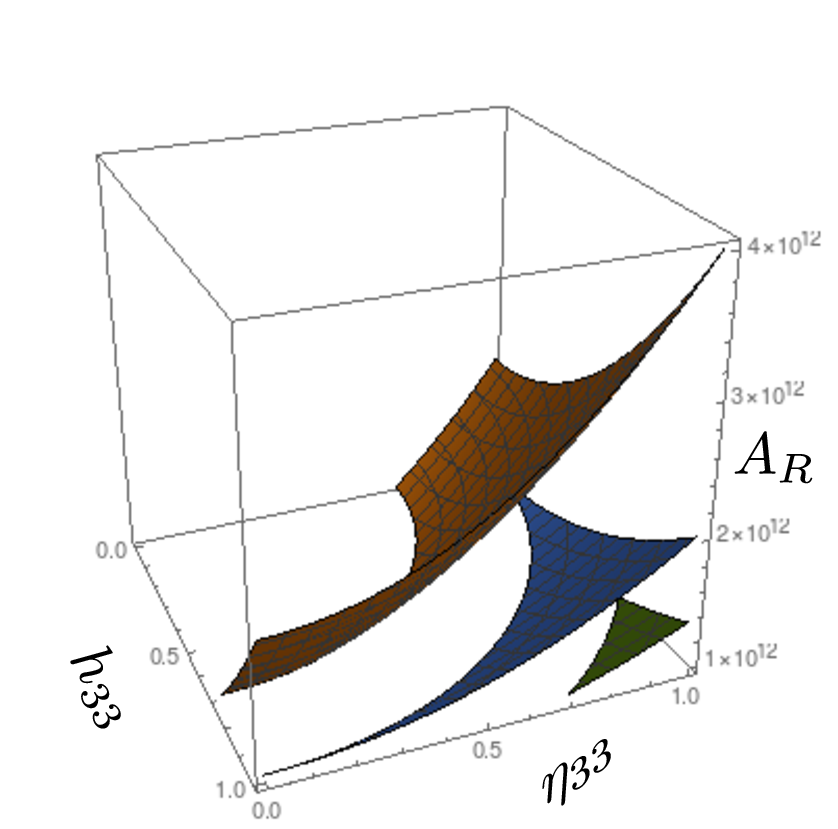}
    \caption{(a)}
  \end{subfigure}
  \hfill
  \begin{subfigure}[b]{0.45\linewidth}
    \includegraphics[width=\linewidth]{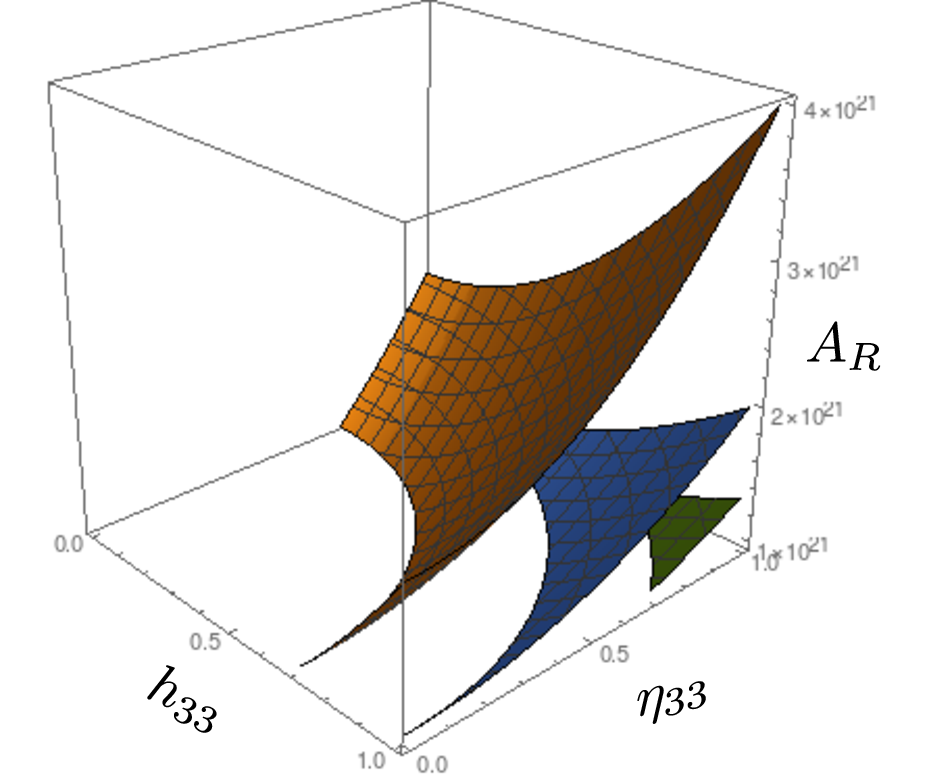}
    \caption{(b)}
  \end{subfigure}
  \caption{The allowed planes of $h_{33}$ and $\eta_{33}$ for $A_R$ in low energy scale regime
  in figure (a) and for high energy scale regime in figure (b)}.
  \label{fig:low-high}
\end{figure}

Taking $h_{33}$ and $\eta_{33}$ in the range $(0-1)$ for two instructive values $A_R$ of one in low scale $\sim 10^{3}$ GeV and other in the high $\sim 10^{12}$ GeV we find the parameter space available for both the hermitian and nonhermitian Yukawa coupling. In figure (\ref{fig:low-high} (a)) and (\ref{fig:low-high} (b)) the allowed planes for $h_{33}$ and $\eta_{33}$
are shown for two different values of $A_R$. But the phases remain unrestricted.
\section{CP violation in lepton sector}
\label{sec:cpv}
The measurement of $CP$ violation in the lepton sector is an interesting
avenue being sought by two current experiments; T2K \cite{Abe:2019vii} and NOvA \cite{Acero:2019ksn}. 
The latter has excluded the absence of $CP$ violation at $95\%$ confidence
whereas the former is consistent with both presence and absence of $CP$ violation. The  $CP$ violation can
be measured in neutrino oscillation experiments by measuring the Dirac
$CP$ phase, $\delta_{\rm CP}$. Moreover, if the neutrinos are Majorana
particles the two Majorana phases introduce some novel features in the leptonic $CP$ violation. The Dirac $CP$ phase leads to $CP$ asymmetry in
oscillation experiments, $P(\nu_\alpha\rightarrow \nu_\beta) \neq P
(\bar{\nu}_\alpha \rightarrow \bar{\nu}_\beta)$ whereas the Majorana
phases do not appear in the oscillation pattern. Nevertheless, Majorana
phases can affect significantly the rates of neutrinoless double beta-decay.

\begin{figure}[h!]
  \centering
  \begin{subfigure}[b]{0.45\linewidth}
    \includegraphics[width=\linewidth]{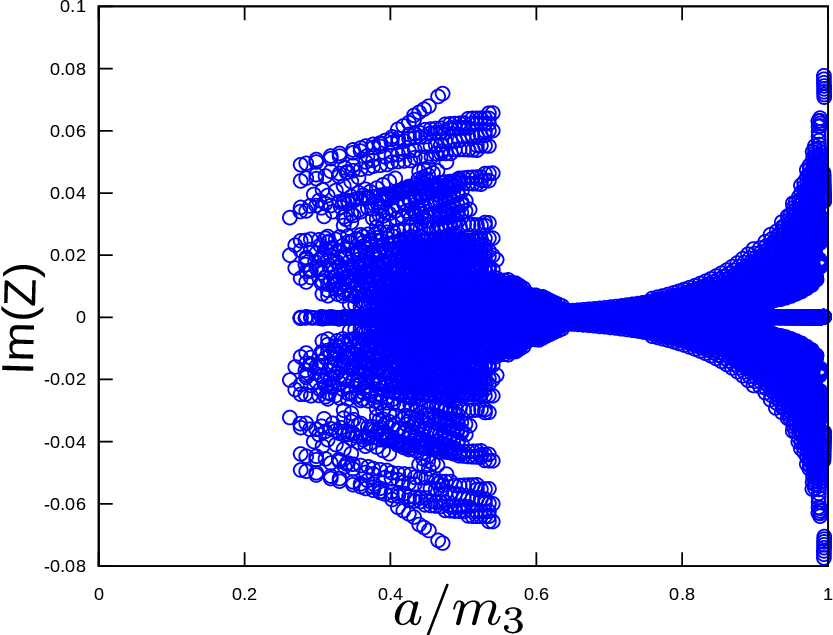}
    \caption{}
     \label{fig:a-z}
  \end{subfigure}
  \hfill 
  \begin{subfigure}[b]{0.45\linewidth}
    \includegraphics[width=\linewidth]{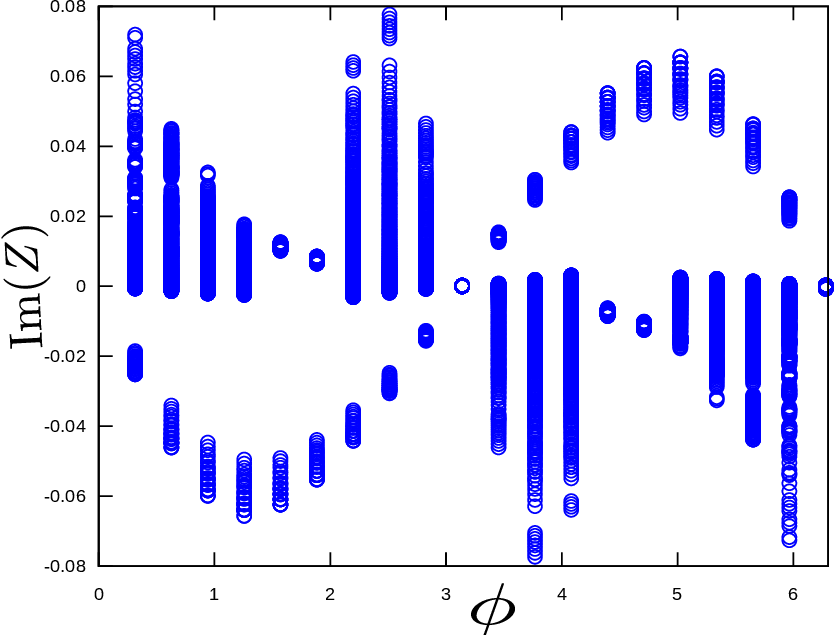}
    \caption{}
     \label{fig:phi-z}
  \end{subfigure}
  \caption{Figure (a) shows the values Im($Z$) vs $a'$. Figure (b) shows the values of  Im($Z$) vs $\phi$ (in radian). }
\end{figure}

The alternative and complementary method to establish $CP$ violation is by 
analyzing the leptonic unitarity triangle similar to the quark sector \cite{AguilarSaavedra:2000vr,Smirnov:2008nh,Dueck:2010fa,Xing:2015wzz}. The link between the leptonic unitarity triangle and neutrino oscillation can be found in \cite{Farzan:2002ct,He:2013rba}. There are a total six
triangles can be constructed using the unitary condition of the lepton flavor
mixing matrix, $U$: three corresponding to the orthogonality of the rows 
(Dirac triangles) and the rest from that of columns (Majorana triangles) given by,
\begin{eqnarray}
\nonumber
  U_{e 1} U_{\mu 1}^* + U_{e 2} U_{\mu 2}^*+ U_{e 3} U_{\mu 3}^* & = & 0, \\
  \nonumber 
U_{\mu 1} U_{\tau 1}^* + U_{\mu 2} U_{\tau 2}^*+ U_{\mu 3} U_{\tau 3}^* & = & 0, \\
 U_{\tau 1} U_{e 1}^* + U_{\tau 2} U_{e 2}^*+ U_{\tau 3} U_{e 3}^* & = & 0. \\
\nonumber
U_{e 1} U_{e 2}^* + U_{\mu 1} U_{\mu 2}^*+ U_{\tau 1} U_{\tau 2}^* & = & 0, \\
\nonumber 
 U_{e 2} U_{e 3}^* + U_{\mu 2} U_{\mu 3}^*+ U_{\tau 2} U_{\tau 3}^* & = & 0, \\
 U_{e 3} U_{e 1}^* + U_{\mu 3} U_{\mu 1}^*+ U_{\tau 3} U_{\tau 1}^* & = & 0.\\
 \nonumber
\end{eqnarray}
All the triangles have common area, $A$ which is proportional to Jarlskog 
invariant, $A=J_{\rm CP}/2$. Absence of $CP$ violation can be inferred
from vanishing $J_{CP}$. The charged-lepton fields can transform under rephasing, 
$l_{{L_j},{R_j}}\rightarrow e^{i\phi_j}l_{{L_j},{R_j}}$. So, the matrix elements of $U$
transform as, $U_{ij}\rightarrow e^{i\phi_j} U_{ij}$. Under rephasing transformation
the Dirac triangles rotate in the complex palne, $U_{ij} U^*_{ik}\rightarrow 
e^{(\phi_j-\phi_k)} U_{ij} U^*_{ik}$. So they have no physical meaning. But the Majorana triangles remain invariant and hence do have physical meaning.
In analogy with the quark sector, the triangle corresponding to the unitary conditions in the first and third column with proper rescaling is shown in NuFit 5.0 \cite{Esteban:2020cvm}. The figure indicates the absence of CP violation would imply a flat triangle i.e., Im$(Z)=0$, where $Z$ is
 \begin{equation}
Z = - \frac{U_{e1} U^*_{e3}}{U_{\mu 1} U^*_{\mu 3}}.
\end{equation}
\begin{figure}[!h]
  \centering
  \begin{subfigure}[b]{0.45\linewidth}
    \includegraphics[width=\linewidth]{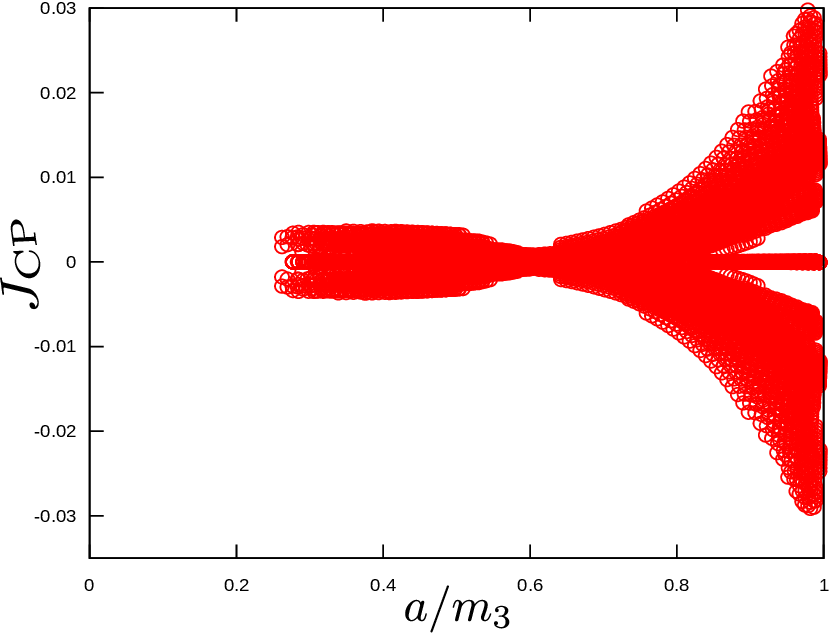}
    \caption{}
     \label{fig:a-jcp}
  \end{subfigure}
  \hfill 
  \begin{subfigure}[b]{0.45\linewidth}
    \includegraphics[width=\linewidth]{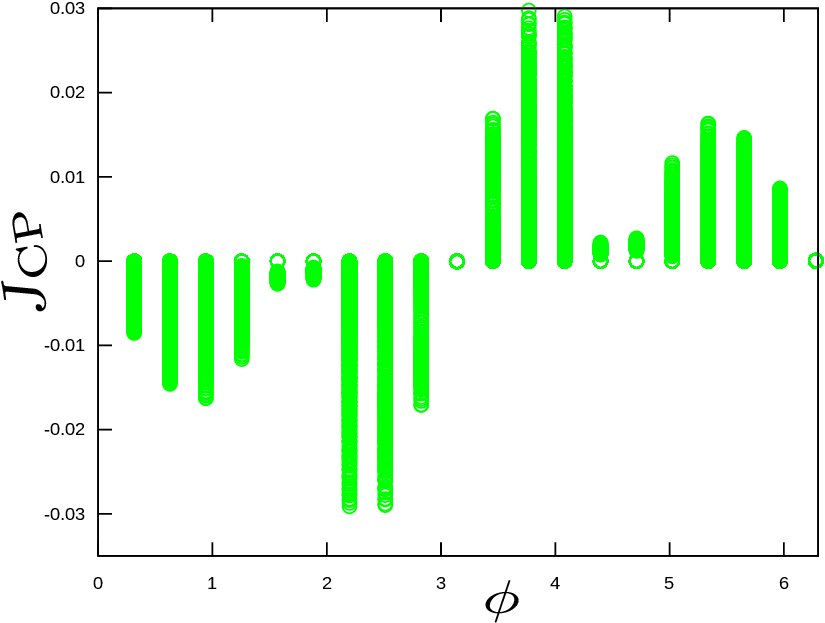}
    \caption{}
     \label{fig:phi-jcp}
  \end{subfigure}
  \caption{Figure (a) shows the range of $J_{\rm CP}$ vs $a'$. Figure (b) shows the values of  $J_{\rm CP}$ vs $\phi$ (in radian). }
\end{figure}
We numerically calculate $J_{\rm CP}$ and ${\rm Im} (Z)$ as
functions of non-Hermitian parameters $a'$ and the phase $\phi$. To examine
the constraints appearing on the parameters from $CP$ violation in
lepton sector the firures in 
 figs.(\ref{fig:a-jcp}), (\ref{fig:phi-jcp}), (\ref{fig:a-z}) and (\ref{fig:phi-z}) can be read together.  It shows the allowed values of the phase $\phi$ which can
 generate $CP$ violation. The $J_{\rm CP}$ value from latest global analysis
 falls in the range, $\sim (-0.035 - 0.04)$ which seems to be achievable in this
 analysis. Further it shows only for specific values of the phase $\phi$, $CP$ violation
 is plausible. Also from the figs.(\ref{fig:phi-jcp}) and (\ref{fig:phi-z}) it is clear
 that $\phi \sim \pi$ should not allowed as it leads to $J_{\rm CP}$ and  ${\rm Im} (Z)$
 to be zero signaling, there is no $CP$ violation. In future the neutrino oscillation
 experiments will quantify the leptonic $CP$ violation more accurately.

{\bf }
\section{Leptogenesis}
\label{sec:lgnsis}
Leptogenesis \cite{Fukugita:1986hr} is an attractive scenario that can
generate baryon asymmetry of the Universe (BAU) throughout of equilibrium decay
of heavy right-handed Majorana neutrinos. Thus it creates a bridge between the seesaw
mechanism and BAU. 
The baryon asymmetry from leptogenesis can be approximated by \cite{Buchmuller:2003gz},
\begin{equation}
 Y_B  \simeq \times 10^{-2} \epsilon_1 \kappa,
\end{equation}
where $k$ is the efficiency factor calculated by solving the Boltzmann equations. $\epsilon_1$ is the
CP-violating decay asymmetry stemming from the interference between tree level and one loop contributions
in the decay of $N_1$, the lightest of the three heavy Majorana neutrinos.
The observations, $Y_B \simeq (6.11 \pm 0.19) \times 10^{-10}$ are
typically compatible with $\epsilon_1\sim (10^{-6} - 10^{-7})$ and $\kappa \sim (10^{-3} - 10^{-2})$.
In the generic realizations of Type-I seesaw 
scenarios the decay asymmetry is given by,
\begin{equation}
 \epsilon_1  \simeq \frac{1}{8 \pi \upsilon^2} \frac{1}{\left(m_D^\dagger m_D\right)_{11}}
 \sum_{j= 2,3} {\rm Im} (m_D^\dagger m_D )^2_{1j} \left( f\left(\frac{M_j^2}{ M_1^2}\right) +
 g\left(\frac{M_j^2}{ M_1^2}\right)  \right),
   \end{equation}
The functions $f$ and $g$ come from vertex \cite{Fukugita:1986hr, Luty:1992un,Plumacher:1996kc}, 
self-energy \cite{Flanz:1994yx, Covi:1996wh, Pilaftsis:1997jf} contributions respectively.
The fraction $M_j^2/M_1^2$ is the ratio of the mass squared of right-handed heavy Majorana neutrinos. 
For a strong hierarchy among the Majorana mass scales $M_1$ and $M_2, M_3$, the ratios 
$M_j^2/M_1^2 \gg 1$ and the functions $f$ and $g$ can be approximated to be
\begin{equation}
 f\left(\frac{M_j^2}{M_1^2}\right)+ g\left(\frac{M_j^2}{M_1^2}\right)  \simeq -3/2 \left(\frac{M_j}{M_1}\right).
\end{equation}
The decay asymmetry, $\epsilon_1$ depends on the Majorana masses and imaginary part of
the combination of Yukawa coupling matrix, $(h_{\nu}^\dagger h_{\nu})^2$. So the Dirac Yukawa coupling, 
$h_\nu$ must be complex for $\epsilon_1 \neq 0$. In generic realizations of seesaw models, $h_{\nu}$ contains most of the
unknown parameters. 

Coming to the non-Hermitian version of Type-I seesaw mechanism the asymmetry produced because of the decay of
the lightest of the three right handed Majorana neutrinos would be,
\begin{equation}
\epsilon_1 \simeq - \frac{3}{16\pi\upsilon^2} \frac{1}{(m_+ m_-)_{11}}
\left( {\rm Im} (m_+ m_-)^2_{12} \frac{M_1}{M_2} + 
{\rm Im} (m_+ m_-)^2_{13} \frac{M_1}{M_3}\right). 
\end{equation}
Using eq.(\ref{eq:four-zero}) the $CP$ asymmetry can be further calculated to be,
\begin{equation}
\epsilon_1 \simeq - \frac{3}{16\pi}
\left[\left( f_1(h_i) \sin \alpha_1 + f_2(h_i) \sin \alpha_2 \right)\frac{M_1}{M_2} + 
\left( f_3(h_i) \sin\beta_1 +f_4(h_i) \sin\beta_2 \right) \frac{M_1}{M_3}\right].
\label{eq:epsilon}
\end{equation}
where $\alpha_1 = \phi_1-\phi_3 , \alpha_2 =\phi_6-\phi_4+\phi_1-\phi_5,
\beta_1= \phi_1 -\phi_5$ and $\beta_2 =\phi_1-\phi_3+\phi_4-\phi_6 $ are different combinations of phases of non-Hermitian Yukawa matrix. The functions
$f_i(h_i)$'s are all quadratic functions of moduli of the non-Hermitian Yukawa matrix. The phases $\alpha_{1,2}(\beta_{1,2})$ are the $CP$ violating phases.
The $CP$ violation parameter depends on the phases and the moduli of the Yukawa matrix and the ratio of right-handed neutrino mass scales. A sizable amount of asymmetry would be
produced for the phase values $\sim (2n+1)\pi/2$.
If the right-handed neutrinos are hierarchical i.e. $M_3>>M_2> M_1$ then the last term will have
negligible contribution as compared to the first two in eq.(\ref{eq:epsilon}).
As discussed in section (\ref{sec:nonhermYukawa}) we are working in the limit where non-Hermitian parameters are of order ${\mathcal O}(1)$, the $CP$ violation of an amount
$(10^{-6} - 10^{-7})$ can be generated if the moduli are of order $0.01$.


\section{Conclusion}
\label{sec:concl}
Non-hermitian extension of Quantum Mechanics has gathered considerable attention in the
recent past. It allows for a wide class of Lagrangians or Hamiltonians for study which might
have been rejected on the basis of hermiticity. In this context, the non-hermitian version
of QED has been pursued \cite{Bender:2005zz}, in particular, its application in neutrino physics \cite{Alexandre:2015kra}. In the usual hermitian version of the SM and in the neutrino sector the Yukawa coupling of neutrinos with the SM leptons appears as $h_{\nu}$ and its hermitian conjugate $h_{\nu}^{\dagger}$ which ensures the mass eigenvalues are real. In the model we have studied the Yukawa couplings in this case are $h_{\pm} = h \pm \eta$, where $h$ and $\eta$ are complex matrices being hermitian and non-hermitian respectively. Here we have calculated the non-Hermitian mass matrix for neutrinos for $3$ generation case in type-I seesaw mechanism and have identified certain constraints which can be helpful to parameterize $h$ and $\eta$. Also, the fine-tuning between the parameters of $h$ and $\eta$ can account for both hierarchical as well as degenerate neutrinos. The requirement of the smallness of neutrino mass also supports a TeV scale of right handed neutrino mass scale which has potential application for low energy scale seesaw.  We consider the seesaw realization of four zero texture of non-Hermitian lepton mass matrices. The mass matrices are diagonalized to get the leptonic mixing matrix and the mixing angles and phases extracted. The values fall well within the range of current experimental data. Using this information we find the allowed plane for nonhermitian Yukawa coupling matrix parameter for both low and high scale of right handed neutrino mass. We also constrain the non-Hermitian parameters using unitarity triangle in the lepton sector. Also, we find the constraint on phases and moduli of non-Hermitian Yukawa matrix to avoid overproduction of the lepton asymmetry to achieve successful baryogenesis through leptogenesis.
The parameter space available for the TeV scale has a potential for lepton flavor violations effects and leptogenesis. Thus non-hermitian version of the SM opens up a new area of study of physics in the neutrino sector. Also the complete classification of texture zeros in the fermion sector must include the non-Hermitian structure as well. The non-hermitian extension of the SM can be the starting point of generating such textures. 

\section{Acknowledgement}
This work is supported by Science and Engineering Research Board (SERB), 
Department of Science and Technology (DST), India grant no. YSS/$2015/000811$.
SM acknowledges the useful discussions with Amit Rai.

\appendix
\section{Elements of orthogonal matrix}

\label{app:elem}
The elements of the orthogonal matrix that diagonalizes the real symmetric matrix
given in eq. (\ref{eq:htilde})  given by $O_{ij} = v_i(m_j^2)$
\begin{eqnarray}
\nonumber
v_1(m^2_i,a,\delta) &=&(b^2+c^2-cd+d^2-m^2_i)(a^2-b|d+a\delta|+b^2-m^2_i)  \\ 
\nonumber
& & - b(c-|d+a\delta|)(-b|d+a\delta|+b^2+c^2+d^2-m^2_i) ,\\
\nonumber
v_2(m^2_i,a,\delta) &=& -m^2_i(a^2+b^2+c^2-cd)+c(a^2+bd)(c-d)\\
\nonumber
& & +b[c(b-c)+m^2_1]|d+a\delta|+m^4_i ,\\
\nonumber
v_3(m^2_i,a,\delta) &=& [c(d-c)-m^2_i](b^2+c^2+d^2-b|d+a\delta|-m^2_i)\\ 
& &  + c(b-d)(b^2+c^2-cd+d^2-m^2_i).
\end{eqnarray}
 \bibliographystyle{apsrev}

\end{document}